# Preventing Malicious Use of Keyloggers : Using Anti-Keyloggers

Alisha Malla
Vellore Institute of Technology
Vellore,TamilNadu,India
alisha.malla2019@vitstudent.ac.in

Jami Gayatri Manjeera
Vellore Institute of Technology
Vellore,TamilNadu,India
gaya3manjeera@gmail.com

Masani Venkata Lakshmi Pravallika
Vellore Institute of Technology
Vellore,TamilNadu,India
pravallikamasani@gmail.com

Prof.Sudha.S
Professor, School of Computer Science
and Engineering
Vellore Institute of Technology
Vellore, Tamil Nadu, India
sudha.s@vit.ac.in

*Abstract*— **Multinational corporations routinely track how its employees use their computers, the internet, or email. There are roughly a thousand devices on the market that enable businesses to monitor their workforce. Observe what their users do online, in their emails, and on their so-called personal laptops while at work.Our team will develop a key-logger for this Journal project that will allow us to capture every keystroke. Additionally, it will enable us to capture the mouse movements and clicks. This will enable us to keep an eye on how a person uses the internet and every other application on his or her "own" computer. Keyloggers, on the other hand, can also be used to steal data in the form of malware or something comparable. An anti-key logger will also be created to address this issue, allowing us to determine whether a key logger is already monitoring the system. We will be able to remain watchful and maintain the security of the data on our system thanks to the anti-keylogger.**

**Problem Statement -** It's challenging to covertly install a hardware keylogger on another person's device. To tackle this issue, We are therefore using a software keylogger that can be remotely installed on a person's PC to resolve this problem. Without the device owner's knowledge, the keylogger would be running in the background.

However, keyloggers can also be used to steal data in the form of malware or other similar threats. To solve this problem, we'll also develop an anti-key logger that will let us know if a key logger is already watching the system. The anti-keylogger will allow us to continue to be vigilant and ensure the security of the data on our system.

## I. INTRODUCTION

A sort of surveillance software called a key-logger (keystroke logging) has the capacity to record each keystroke made on a machine once it has been installed. Therecordings are kept in a typically encrypted log file.

The maiority of contemporary keyloggers are thought to be trustworthy software or hardware that may be purchased on the open market.Keyloggers may be used ina variety of situations that are both legal and suitable,according to developers and distributors.

Nowadays, privacy of our information matters more than anything. Our analyses suggest that using Anti-Keylogger in addition to antivirus and firewall to ensure complete protection against data leaks and online risks. With the aid of anti-virus, firewall, and anti-keylogger, viruses will be eliminated, unwanted Internet connections will all be cutoff, and spyware will all be banned.

Antikeylogger is a form of software to detect a keylogger on the machine .It is created in a way that it can identify any keylogger that may be present in the system. It is able to prevent any keylogger activity from happening. It is made with the intention of stopping keylogger activity. It shields the user from keyloggers intercepting their keystrokes. Therefore, the user's sensitive or private information can be protected from the attacker utilising Antikeylogger.Technically anti-key loggers are used to protect sensitive information such as passwords, confidential information from detection of key loggers.

Overall, the main technical motive for using an anti-key logger is to protect sensitive information, prevent unauthorized access to your computer, and improve the overall security of your system.Frequently, such software will also have the ability to remove or at the very least disable any hidden keystroke loggers on a computer.

*Types of Antikeyloggers :*

1. **Signature based Anti-Key-Logger**: This kind of program has a signature base, which is tactical data that aids in specifically identifying a keylogger. The list includes as many well- known keyloggers as is practical. Some merchants make an attempt to make an updated listing available for customers to download. This software analyses the contents of the hard drive, item by item, against the list each time a "Systern Scan" is run, looking for any matches.

2. **Heuristic analysis based Anti-Key-Logger:** Despite offering stronger keylogging protection than signature-based anti-keyloggers, this technology is not without its flaws. One of them is the fact that non-keyloggers are also blocked by this kind of software. Numerous "non-harmfull" software components, whether they are a component of the OS or genuine apps, use keylogger-specific operations that could result in a false positive. Most non- signature-based keyloggers include the option for the user to manually select which modules to block or unblock, but this might be challenging for non - technical users who are unable to distinguish between good and harmful modules

## II. LITERATURE SURVEY

### A. REVIEW ON VARIOUS SCHEMES

| (Title, Year, Authors) | Methodology or Techniques used (Mention specific algorithms or recent technologie) | Advantages | Issues |
|---|---|---|---|
| 1. Survey of keylogger Technologies Dr. Anas Bilal | They used 3 main methods in this research paper: First method, the operating system-based Windows Keyboard Hook technique offers several functions to the Hook-based keyloggers to keep an eye on the keyboard. The second method uses a table that contains the status of 256 virtual keys and is called the Keyboard State Table method. | Allow for no false negatives when the keylogging behavior is triggered within the window of observation and can also be used in large-scale malware analvsis and classification. | Inconsistent time periods have no quantitative analysis. |
| 2. Analysis and Implementation of Decipherments of KeyLogger, Parth Mananbhai Patel | In this they have developed a detection method for Type I and Type II user-space keyloggers. | The option to artificially insert carefully selected keystroke patterns was added to the model in this research study, and the issue of selecting the ideal input pattern to increase our detection rate was examined | There is no quantitative analysis for irregular time intervals. |
| 3. Keyloggers: A Malicious Attack, Dr.c.u marani rajrishi sengupta | In this Survey different examination regions which spread convention confirmations representation of keylogging assaults | It has high detection rate and a low false alarm rate. | All apps that hook the system, even those that are genuine would be flagged as malicious. |
| 4. Anti-Hacking Mechanism for Keylogger using Blackbox Detection, Saiganesan N, Dheenadhayalan. A, Arulmani. M, Suresh K | In this study, they carefully correlated the input and output to replicate the behaviour of a keylogger. | It can quickly locate any suspicion processes or files, whether they are visible or not at any application level. | This method takes extensive processing, and the rate of false positives is really high |
| 5. Survey on keystroke logging Attacks, Kavya .C, Suganya.R | This study paper aims to provide an understanding of how keyloggers operate, various password attacks, and preventive and detection techniques to lessen and prevent keylogging attacks. | With this method, a keylogger records every keystroke the user makes and saves the information in a file on the hard drive. | False positive rate is very high. |
| 6. Detecting Software Keyloggers with Dendritic Cell Algorithm., J. Fu atel. | A hook programme is used by the Dendritic Cell Algorithm to track API calls produced by active processes. Five signals are used in the host to define the system's state. | This approach has a high detection rate and a low false alarm rate, and it can distinguish between the active keylogger activity and the regular processes. | Keylogger behaviour is the same as that of programmes that hook system message execution. The system would identify all normal programmes that attach it as malicious. |
| 7. Cyber Crime Security implementation using Hardware based Anti-keylogger Dr.D.B.V Jagannadham, Mr.D.Ajit Varma | In this paper, we show how to use an external anti-key-logging device to find a key-logger. It recognises the hardware keylogger and notifies the user of the danger. | Developed a hardware-based anti-keylogger that is especially made to detect any keystroke-capturing device | This approach requires a lot of processing, and false positive is very high. |
| 8. System Monitoring and Security Using Keylogger, Preeti Tuli, Priyanka Sahu | This essay makes an effort to define "Informational Flow" on the internet and computer usage as concretely as possible, including its scope, the fundamental ideas involved, and the factors influencing its uptake. | Software that can both monitor each keystroke and action made on a PC. | This will be more harmful if a keylogger and social engineering skills are combined |

| 9. Detecting Kernel Level Keyloggers Through Dynamic Taint Analysis Mohd Aizaini Maarof ,FuadMire Hassan and Mohamed Muse Abshir | Keyloggers at the kernel levelcan be found using dynamic taint analysis and host-based intrusion detection. | In this essay a general review of keylogger programs.This paper uses black berry as a casy study. | It's vital to integrate Vm scopes approaches. |
|---|---|---|---|
| 10. Design and Implementation of Detection of KeyLoggers | In order to unmistakably identify the keylogger among all the active processes, they have presented a new detection method that observes the behaviour of the keylogger in output after simulating designed keystroke sequences in input. . | The prototype system was successfully tested against the most popular free key loggers, with no reported false positives or false negatives. The lengthy discussion of potential attacks on detection techniques is offset by the technique's simplicity in use. | The system would identify all normal programmes that attach it as malicious. |

**Techniques of keylogger detection and Antikeylogger Detection :**

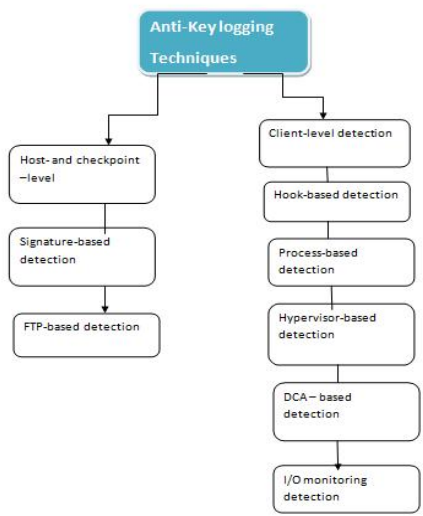

Anti-keylogger Techniques(Fig 1)

### III. REQUIREMENTS

For the keylogger2 and anti-keylogger, the system needs to have installed java compiler along with Native libraries. The best way to ensure this is to install Microsoft Visual Studio code on the system, as it might help in installing any of the other missing libraries that might be required to run the program.

### IV. METHODOLOGY

The key-logger is built in java language as J Native libraries from the Apache common providers will be utilized . We choosen java language because this it is faster when compared to the python code and can also monitor cursor movements, cursor location and key-presses. J-Native Hook uses platform-dependent native code in conjunction with Java's native interface to build low-level system-wide hooks and send events to the application. It runs hidden in the background and captures keystrokes.

General representation of Keylogger :

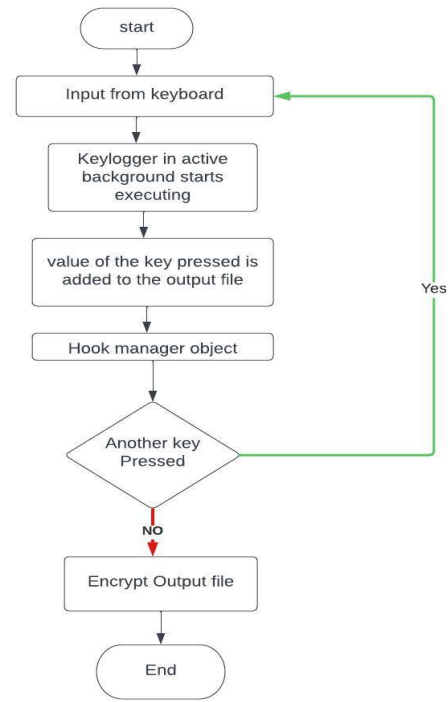

Step 1 : Start

Step 2 : The input values are all the keystroke. (all the keys pressed from the keyboard)

Step 3 : After a key is pressed on the keyboard, the execution of the key-logger program active in the background starts.

Step 4 : Now, the value of the key pressed, i.e., alphabet or number or special character will be written in the output file.

Step 5 : The program will create a hook manager object and set it for the keystrokes and info to follow.

Step 6 : After this, the key logger will save all the data on the log file as an output file.

Step 7 : End

The output file will be encrypted using the hill cypher algorithm as it helps in monitoring the inputs and as protects the data from outsider.

This uses a client and server socket programming concept where the data is monitored from the client and sent to the server. In our example we will be using both server and client on the same system (local host) but the server can be changed accordingly.The keystrokes are recorded on the server side and stored in a local file

Working of Anti-keylogger :

The Anti-keylogger is built in java as it will be working on the concept of java file handling. Java file handling enables the file class from the java.io package, allows us to work with different formats of files. In order to use the File class, you need to create an object of the class and specify the filename or directory name

A. Architecture:

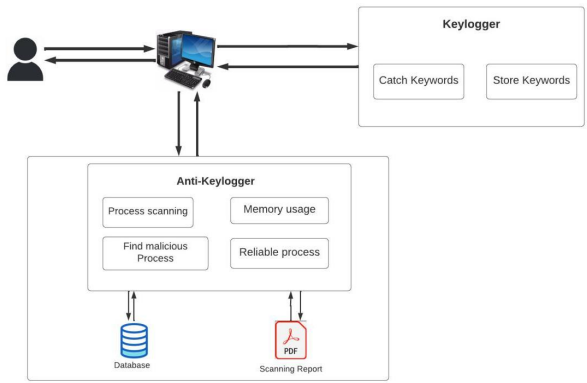

The signature file of over 1000 headers have been taken from Git-hub whose link has been added below in the references.

Constructors and Methods sare :

Public Detector(String path) : Initializes the Detector object with the given path and initializes the file, affectedFiles, resultFile, signatures, and errorFile fields. It also reads the signatures from the signatures file and stores them in the globalMd5 field.

Public void detect() : a public method that serves as a wrapper function to call the detect method with the file field as a parameter.

Private void detect(File file) : a private recursive method that detects keyloggers based on their signatures. It checks if the file parameter is not null and if it's a directory or a file. If it's a directory, it recursively calls the detect method with each file in the directory as a parameter. If it's a file, it calculates its SHA1 hash and checks if it matches any of the known keyloggers' signatures stored in the globalMd5 field. If there is a match, it writes the path of the affected file to the affectedFiles file.

Private void writeResult(String result) throws IOException: A private method that writes the result parameter to the file parameter.The code also imports classes from the Apache Commons Codec and IO libraries to calculate the SHA1 hash of files and read the contents of files, respectively.

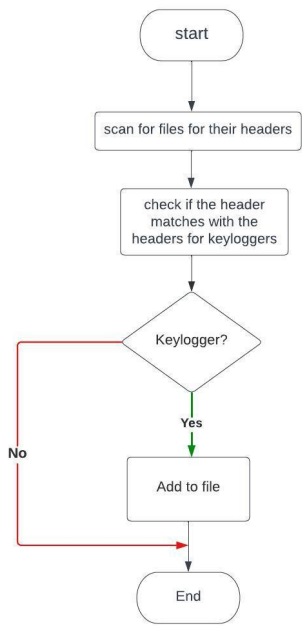

Step 1: Start

Step 2: Scans for files

Step 3: Compare the headers of these files with the pre-given file headers obtained from the internet of over 1000 various anti-keylogger headers.

Step 4: If a file is detected matching with the given header, its name is added to a file.

Step 5: The user checks the output file, and decides the action to be taken on the file names mentioned.

Step 6: End

The code detects keyloggers based on signatures. It is built in java language as JNative libraries from the apache common providers will be utilised. This is done because this library is faster when compared to the python code and can also monitor cursor movements, cursor location and keypresses.

V. IMPLEMENTATION & RESULTS

The keylogger implementation module involves creating a program that captures keystrokes and listens to mouse events. The purpose of this code is to explore the potential for malicious use, as it can record user input on both keyboard and mouse. The module is composed of two source files, namely Client.java and Server.java. The output of this module includes the logs of captured keystrokes and mouse events.

We developed a detection mechanism to identify the presence of keyloggers. The mechanism includes a detector that scans for known signatures of malicious software in order to detect any such software running on the system. The keylogger detection module consists of two source files, Detector.java and Main.java. The input file required for this module is signatures.txt, while the output files generated by the module include affected.txt, result.txt, and errors.txt.

To activate the keylogger these are the steps :

Keylogger 1:
The requirements to run this keylogger are given below -
1. Python 3 must be installed on the system
2. Make sure to have "pynput" package installed (Can be done by running the following command in terminal: "pip install pynput").
After the execution of the program the log file start storing all the keys pressed To run the program as a hidden agent, save the code file with the extension
"pyw"

Keylogger 2:
To execute the keylogger, open command prompt in the folder where all the java files are saved along with jar files and bat files.
Step 1: Execute the server.bat file in command prompt.
Step 2: Open another command prompt
Step 3: Execute client.jar file
The output of the keylogger will be visible in the server terminal and getting saved in the output file.

Anti-keylogger :

To execute the anti-keylogger, open command prompt in the folder where all the java files are saved along with jar files and bat files.

Step 1: Execute the detect.bat file.
Step 2: Follow the instructions on the screen.

After the correct execution of the instructions the suspected files are added to the output file and the errors faced the error file.

This will launch the detector automatically. Follow the on-screen instructions to proceed.

If you are using a Mac or Linux system, open the terminal or command line and enter the following commands to start the keylogger detector:

    $ cd <folder where kelogger detector files reside>
    $ java -jar keydetect.jar

| Metric | Keylogger 1 | Keylogger 2 |
|---|---|---|
| Language | Python | Java |
| Library used | Pynput | Jnative |
| Operating System | Windows | Windows |
| Hidden | If the file extension is pyw | NO |
| Records keyboard keystrokes | Yes | Yes |
| Records mouse location and clicks | No | Yes |
| Output File location | Needs to retrieve from the target system originally | Will be shared to the specified email in the program |

## VI. CONCLUSION

As we can see from the above implementation that we have created a keylogger in java language after implementing a python keylogger for windows. The table below gives more information regarding the similarity and differences between the two keylogger